%% file: limitedLifetimes.tex
\documentclass[a4paper, 10pt, onecolumn]{article}

\usepackage{fancyhdr}

\newcommand{\comment}[1]{}
\newcommand{\tit}{Reduction of the Statistical Power Per Event Due to
                  Upper Lifetime Cuts
		  in Lifetime
		  Measurements
}
\newcommand{\me}{Jonas Rademacker}

\usepackage[dvips]{color}
\usepackage[dvips,bookmarks=true,bookmarksopen=true,bookmarksnumbered=true,a4paper=true,nesting=false,colorlinks=false]{hyperref}
\usepackage[dvips]{graphicx}

\hypersetup{%
pdftitle = {\tit}
pdfauthor = {\me},
pdfkeywords = {lifetime, fit, trigger, impact parameter, bias}
}

\input{units}

\input{btodst_commands}

\fontsize{12}{24}
\newcommand{\Figref}[1]{Figure \ref{#1}}
\newcommand{\figref}[1]{Figure \ref{#1}}
\newcommand{\Eqref}[1]{Equation \ref{#1}}
\newcommand{\eqref}[1]{Equation \ref{#1}}

\newcommand{\secref}[1]{Section \ref{#1}}

\newcommand{\fig}{./}

\newcommand{\st}{\ensuremath{\mbox{}^{\mathrm{st}}}}
\newcommand{\nd}{\ensuremath{\mbox{}^{\mathrm{nd}}}}

\newcommand{\dbyd}[2]{\ensuremath{\frac{\mathsf{d}#1}{\mathsf{d}#2}}}
\newcommand{\tmin}{\ensuremath{t_{\mathrm{min}}}}
\newcommand{\tmax}{\ensuremath{t_{\mathrm{max}}}}
\newcommand{\tmini}{\ensuremath{t_{\mathrm{min}\, i}}}
\newcommand{\tmaxi}{\ensuremath{t_{\mathrm{max}\, i}}}
\newcommand{\func}[1]{\ensuremath{\mathrm{#1}\!}}
\newcommand{\Lik}{\ensuremath{\mathcal{L}}}
\newcommand{\logLik}{\ensuremath{\log\Lik}}

\newcommand{\gauss}[2]{\ensuremath{ \frac{1}{\sqrt{2\pi}} e^{-\frac{{#1}^2}{2{#2}^2}}   }}

\newcommand{\Freq}{\ensuremath{\mathrm{F}}}

\newcommand{\labelpawx}[6]{heinz}

\begin{document}

\title{\tit}
\author{
 \renewcommand{\thefootnote}{\arabic{footnote}}
\me\footnotemark[1]
\\ \textit{University of Bristol}
}
\date{}
%
\maketitle
\fancyfoot{}
  \renewcommand\headrulewidth{0pt}
  \renewcommand\footrulewidth{0pt}
\setcounter{page}{0}
\thispagestyle{fancy}
\vfill
\begin{abstract}
 A cut on the maximum lifetime in a lifetime fit does not only reduce the
 number of events, but it also, in some circumstances dramatically,
 decreases the statistical significance of each event. The upper
 impact parameter cut in the hadronic B trigger at CDF \cite{CDF}
 \cite{SVT} \cite{XFT}, which is due to technical limitations, has the
 same effect. In this note we describe and quantify the consequences
 of such a cut on lifetime measurements. We find that even moderate
 upper lifetime cuts, leaving event numbers nearly unchanged, can
 dramatically increase the statistical uncertainty of the fit result.
\vspace{1ex}\\
Keywords: {\it lifetime fit; lifetime cuts; impact parameter
  cuts; lifetime bias; hadronic B trigger; statistical power
  per event; CDF; B Physics}
\\
PACs: { 21.10.Tg; 02.50.-r; 14.40.Nd}
\end{abstract}
\footnotetext[1]{Jonas.Rademacker@bristol.ac.uk}
%
\input{theMeat}

\end{document}

%% file: units.tex
\newcommand{\un}[2]{\ensuremath{\mathrm{#1 \, #2}}}


%% file: btodst_commands.tex
\newcommand{\prt}[1]{\ensuremath{{\rm #1}}}

%% file: theMeat.tex
\section{Introduction}
\label{sec:intro}
 In this note we discuss the impact of an upper lifetime cut on the
 precision of a lifetime measurement. We find that even an upper
 lifetime cut that loses only a small fraction of the data can have
 dramatic consequences on the precision of the lifetime fit, due to a
 loss of statistical power of those events that pass the cut. The
 small loss of events due to a moderate upper lifetime cut is
 accompanied by a large loss of information, because not only a few
 events outside the allowed time window are lost, but also the
 information that there were only a few. This can have dramatic
 effects on the precision of the measurement. As shown below, an upper
 lifetime cut that loses \un{14}{\%} of the data reduces the
 statistical significance per event by \un{72}{\%}, so the combined
 effect on the statistical precision of the lifetime measurement is
 equivalent to losing \un{76}{\%} of the data.

\begin{figure}
\begin{center}
\caption[Impact parameter cut translates to lifetime cut]{Given the
 3-momenta of all particles in the decay, the cut on the Impact
 parameter of the decay products translates directly into a cut on the
 decaylength and hence on the
 lifetime of the primary particle. For clarity, the figure only
 illustrates the effect of an impact parameter cut on one of the decay
 products (the one going straight
 upwards).\label{fig:ip_to_decaylength}}
\includegraphics[width=0.33\columnwidth]{\fig/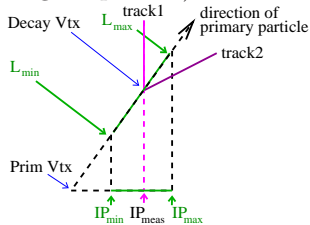}
\end{center}
\end{figure}

 Such a cut on the maximum lifetime is for example implicitly applied
 in the hadronic trigger sample at CDF \cite{CDF}, where the trigger
 requires two tracks with a minimum impact parameter of
 \un{100-120}{\mu m} (depending on the exact configuration) and a
 maximum impact parameter of \un{1000}{\mu m} \cite{SVT}
 \cite{XFT}.  These impact parameter cuts translate into upper and
 lower lifetime cuts, which differ event by event. This is illustrated
 in \figref{fig:ip_to_decaylength}.

\section{Likelihood with lifetime cuts}
 We can write the probability to find an event with decay time $t$,
 given that it passed the trigger cuts, as:
\begin{eqnarray}
\label{eq:probBasic}
\func{P}\left(t| t\in\left[\tmin,\tmax\right]\right) 
   &=& \frac{ \frac{1}{\tau} e^{\frac{-t}{\tau}}
            }{
              \int\limits_{\tmin}^{\tmax}\frac{1}{\tau}e^{\frac{-t^\prime}{\tau}} dt^\prime
            } 
    =  \frac{ \frac{1}{\tau} e^{\frac{-t}{\tau}}
            }{
               e^{\frac{-\tmin}{\tau}} - e^{\frac{-\tmax}{\tau}}
            }
\end{eqnarray}
 where we ignore the effect of measurement errors.
 The total log-likelihood function for a set of $N$
 ``ideal'' two-body decays (no measurement uncertainties, background,
 etc) is given by:
\\\parbox{\columnwidth}{
\begin{eqnarray}
\label{eq:logBasic}
 \logLik&=&  - N\log\left(\tau\right) \nonumber \\
        & &  - \sum\limits_{i=1}^{N}
             \left( \frac{t_i}{\tau} +
                   \log\left( 
                      e^{-\tmini/\tau} - e^{-\tmaxi/\tau}
                   \right)
             \right) \nonumber \\
        & &
\end{eqnarray}
} where the index $i$ labels the event, each of which has its measured
 decay time $t_i$ and minimum and maximum lifetime cuts of \tmini\ and
 \tmaxi.

 Note that the only difference compared to the the likelihood function without
 lifetime or impact parameter cuts is the term:
\begin{equation}
\label{eq:basicCorrection}
     \logLik_{\mathrm{ip}} =
              - \sum\limits_{i=1}^{N}
                \log\left( 
                   e^{-\tmini/\tau} - e^{-\tmaxi/\tau}
                \right)
\end{equation}

Including Gaussian event-by-event measurement errors, the PDF for an
event with lifetime cuts $t_{\mathrm{min}}$ and $t_{\mathrm{max}}$,
and the measured lifetime $t_0$, is given by:
\begin{eqnarray}
\label{eq:likWithErrors}
P   &=& 
   \frac{
     \int\limits_{0}^{\infty}
       \frac{1}{\tau} e^{\frac{-t}{\tau}} \gauss{\left(t-t_0\right)}{\sigma_t} 
     \, dt
   }{
     \int\limits_{t_{\mathrm{min\; i}}}^{t_{\mathrm{max\; i}}}
       \int\limits_{0}^{\infty}
         \frac{1}{\tau} e^{\frac{-t}{\tau}} \gauss{\left(t-t_0\right)}{\sigma_t} 
       \, dt
     \, dt_0
   }\nonumber\\
   &=& 
   \frac{
           \frac{1}{\tau}e^{\frac{-t_0}{\tau} + \frac{1}{2} \frac{\sigma^2}{\tau^2}
}
           \Freq\left( \frac{t_0}{\sigma} - \frac{\sigma}{\tau}\right)
   }{
    \left[
           - e^{\frac{-t}{\tau} + \frac{1}{2} \frac{\sigma^2}{\tau^2}}
           \Freq\left( \frac{t}{\sigma} - \frac{\sigma}{\tau}\right)
           + \Freq\left(\frac{t}{\sigma}\right)
    \right]_{t=t_{\mathrm{min\; i}}}^{t=t_{\mathrm{max\; i}}}
   }
\end{eqnarray}
with 
\begin{equation}
 \Freq(x) \equiv \frac{1}{\sqrt{2\pi}}\int\limits_{-\infty}^{x} e^{\frac{-y^2}{2}} \,dy
\end{equation}

\section{Calculating the Uncertainty of the Fit Result}
The variance on the fit result can be estimated as the negative
inverse of the 2\nd\ derivative of the likelihood function, evaluated
at the lifetime $\tau$ that maximises the likelihood:
\begin{equation}
\label{eq:varianceFromLogLik}
 \sigma^2 = V = - 1 \Bigg/ \dbyd{^2\logLik}{\tau^2}\Big|_{\tau = \tau_{\mathrm{fit}}}
\end{equation}
 This formula assumes that near its minimum, the Likelihood curve can
 be approximated by a Gaussian. While this approximation is not
 perfect for lifetime fits, that have asymmetric errors, it is still
 very good as long as the event samples are sufficiently large, as
 shown in \secref{sec:MC}.

 For simplicity, we ignore event-by-event measurement errors in our
 subsequent calculations. We will show in \secref{sec:MC} that,
 for the purpose of estimating the error on the fit result, this
 provides a very good approximation for the case of B lifetimes
 measured at CDF where the event-by-event measurement uncertainties
 are much smaller than the lifetime to be measured.  The 1\st\
 derivative of the likelihood function defined in \eqref{eq:logBasic}
 is:
\begin{eqnarray}
 \dbyd{\logLik}{\tau} &=& 
         \frac{1}{\tau^2}
         \left(
           -N\tau 
           + \sum\limits_{i=1}^{N}\left(t_i -\tmini\right)
           + \sum\limits_{i=1}^{N}
                 \frac{\Delta t_i}{e^{\Delta t_i/\tau} -1}
               \right)
\\ & &
  \mbox{with}\;\; \Delta t_i \equiv \tmaxi-\tmini \nonumber
\end{eqnarray}
 where we introduced $\Delta t_i = \tmaxi - \tmini$, the width of the
 time interval to which the $i-\mathrm{th}$ event is confined due to
 impact parameter, decay distance, or direct lifetime cuts. The 2\nd\
 derivative is:
\begin{eqnarray}
\label{eq:secondDerivative}
 \dbyd{^2\logLik}{\tau^2} &=& 
   -\frac{2}{\tau} \dbyd{\logLik}{\tau}
   - \frac{1}{\tau^2}
     \left(N - \sum\limits_{i=1}^{N}
                 \left(
                   \frac{ \frac{1}{2} \Delta t_i / \tau }{
                          \sinh\left(\frac{1}{2} \Delta t_i / \tau
                               \right)
                        }
                 \right)^2
     \right)
\nonumber\\
   &=&
   -\frac{2}{\tau} \dbyd{\logLik}{\tau}
   - \frac{N}{\tau^2}
     \left(1 -   \left<\left(
                   \frac{ \frac{1}{2} \Delta t / \tau }{
                          \sinh\left(\frac{1}{2} \Delta t / \tau
                               \right)
                        }
                 \right)^2\right>
     \right)
\end{eqnarray}
 where the angle bracket indicate taking the mean of the expression
 inside over all events. At the value of $\tau$ that maximises the
 likelihood, the first term of \eqref{eq:secondDerivative} vanishes,
 and the variance is given by:
\begin{equation}
\label{eq:varianceCalculated}
 \sigma^2 = 
  \frac{\tau^2}{
     N -
      \sum\limits_{i=1}^{N}
      \left(
      \frac
          { \frac{1}{2}
              \Delta t_i/\tau
          }{
            \sinh\left(\frac{1}{2} \Delta t_i /\tau\right)
          }
      \right)^2
  }
  =
  \frac{\tau^2}{
    N 
  }
  \cdot
  \frac{1}{
     1 -
      \left<
      \left(
      \frac
          { \frac{1}{2}
              \Delta t/\tau
          }{
            \sinh\left(\frac{1}{2} \Delta t /\tau\right)
          }
      \right)^2
      \right>
  }
\end{equation}
 Note that the lower lifetime cut by itself does not have any impact
 on the statistical precision (apart from changing the number of
 events), it is the width of the time interval defined by the cuts,
 that matters; it is therefore the presence of an upper lifetime or
 impact parameter cut that affects the statistical precision per
 event.

\section{Statistical Power Per Event}
The right hand side of \eqref{eq:varianceCalculated} can be
separated into two factors:
\begin{itemize}
 \item The variance in the absence of any upper lifetime cut:
 \begin{equation}
   \label{eq:varNoCuts}
    \frac{\tau^2}{N}.
 \end{equation}
 \item The correction factor due to the upper lifetime cut:
 \begin{equation}
   \label{eq:cutEffect}
   \frac{1}{
     1 -
      \left<
      \left(
      \frac
          {\Delta t/\tau
          }{
            \sinh\left(\frac{1}{2} \Delta t /\tau\right)
          }
      \right)^2
      \right>
     }
 \end{equation}
\end{itemize}
 So the change in statistical precision per event due to an upper
 lifetime cut is accounted for by making the following replacement for
 $N$:
\begin{equation}
 N \longrightarrow
 N\cdot 
   \left(
     1 -
      \left<
      \left(
      \frac
          { \frac{1}{2}
              \Delta t/\tau
          }{
            \sinh\left(\frac{1}{2} \Delta t /\tau\right)
          }
      \right)^2
      \right>
    \right).
\end{equation}
 It makes therefore sense to define the \emph{statistical power per
 event}, $\mathcal{P}$, as
\begin{equation}
\label{eq:defStatPower}
 \mathcal{P} \equiv       1 -
      \left<
      \left(
      \frac
          { \frac{1}{2}
              \Delta t/\tau
          }{
            \sinh\left(\frac{1}{2} \Delta t /\tau\right)
          }
      \right)^2
      \right>.
\end{equation}
 $\mathcal{P}$ is $1$ for events without upper lifetime cuts, and $<1$
 otherwise. It is defined such that $N$ events with an upper lifetime
 cut (where $N$ is the number of events after the cut has been
 applied) are statistically equivalent to \( N \cdot \mathcal{P}\)
 events without an upper lifetime cut.
\begin{figure}
\caption{Statistical Power per Event as a function of $\Delta
  t/\tau$.\label{fig:statPowerPerEvent}}
\begin{tabular}{cc}
\input{Fig_statPower}
&
\parbox{0.39\textwidth}{
$\mathcal{P}$ is defined as
\[
 \mathcal{P} \equiv       1 -
      \left<
      \left(
      \frac
          { \frac{1}{2}
              \Delta t/\tau
          }{
            \sinh\left(\frac{1}{2} \Delta t /\tau\right)
          }
      \right)^2
      \right>,
\]
  where $\Delta t$ is the width of the time window defined by the
  lifetime cuts, $\tau$ is the lifetime to be measured.
\vfill\mbox{}
}
\end{tabular}
\end{figure}
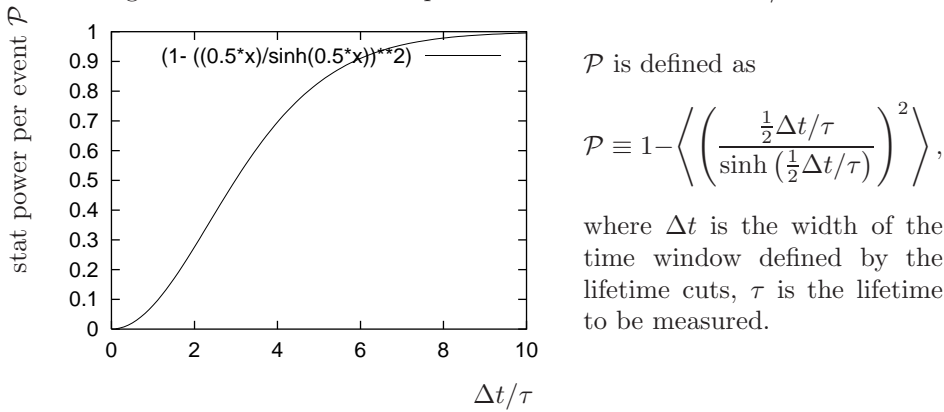
  \Figref{fig:statPowerPerEvent} shows the statistical power per event
  as a function of $\Delta t/\tau$, the time interval defined by the
  cuts divided by the lifetime. Upper lifetime cuts that seem harmless
  at first sight, because they have a small effect on the number of
  events, can have a dramatic impact on the statistical precision of
  the lifetime measurement due the reduction in statistical power per
  event. For example an upper lifetime cut leaving a time interval
  that is twice as wide as the mean lifetime to be measured (\(\Delta
  t /\tau = 2\)) retains $(1-e^{-2})=86\%$ of the events, but the
  statistical power per event is reduced to $28\%$. The combined
  effect is equivalent to losing $100\% - 86\% \cdot 28\% = 76\%$ of
  the unbiased sample before the cut, rather than the naively expected
  $14\%$.

 The CDF hadronic B trigger requires two tracks with impact parameters
 between \un{100}{\mu m} and \un{1000}{\mu m}, which translate to
 different upper and lower lifetime cuts for each event, typically
 yielding a $\Delta t$ between $1$ and $3$ times the B lifetime (this
 is an approximate number from studies of \prt{B^0 \to \pi \pi}
 decays). So each event in that sample is, for the purpose of lifetime
 measurements, only worth about $30\%$ of an unbiased event. Note that
 this is true for lifetime measurements, only, and not for asymmetries
 or oscillation measurements, where it is the oscillation period that
 determines the scale $\Delta t$ needs to be compared to, rather than
 the mean lifetime.

\section{Monte Carlo Studies}
\label{sec:MC}

 In order test how good the error estimate in
 \eqref{eq:varianceCalculated} is, given that this simplified formula
 ignores measurement errors and assumes the likelihood to be Gaussian
 near its minimum, it is compared to the error estimate from a MINUIT
 fit using a likelihood function that includes event by event
 errors. The errors are calculated using the MINOS algorithm within
 MINUIT. The fit is performed on simulated \prt{B^0 \to \pi\pi} events
 at CDF. The Monte Carlo simulation includes a detailed description of
 the CDF detector, including the hadronic B trigger, that requires two
 tracks with impact parameters between \un{100}{\mu m} and
 \un{1000}{\mu m}. This requirement translates into upper and lower
 lifetime cuts, which differ event by event. The likelihood function
 used to fit the simulated data includes the trigger effects, and the
 event-by-event uncertainties in the lifetime measurement, as given in
 \eqref{eq:likWithErrors}. It is in the calculation of the acceptance
 limits where further detector effects are taken into account, in
 particular the difference between the fast-reconstructed online
 impact parameters used by the trigger, and the more precise offline
 reconstruction used in the actual lifetime measurement.
\begin{figure}
\caption{Fit to $\sim 13k$ MC-generated signal events. The shaping of the
  distribution due to the trigger manifests itself as a clear
  deviation from a straight line in this logarithmic plot. The fit
  describes the data well with a $\chi^2/\mathrm{dof} = 0.8$. The
  result of \un{c\tau=459.1^{+7.3}_{-7.1}}{\mu m} is in good agreement
  with the input value of \un{462}{\mu m}. The numerical error
  estimate of $\mbox{}^{+7.3}_{-7.1} \mu m$, calculated using the
  MINOS algorithm in MINUIT, agrees well with analytical value of
  \un{\pm 7.2}{\mu m} from \eqref{eq:varianceCalculated} derived
  here.
  \label{fig:fit}
}
\input{Fig_lifeFit}
\end{figure}
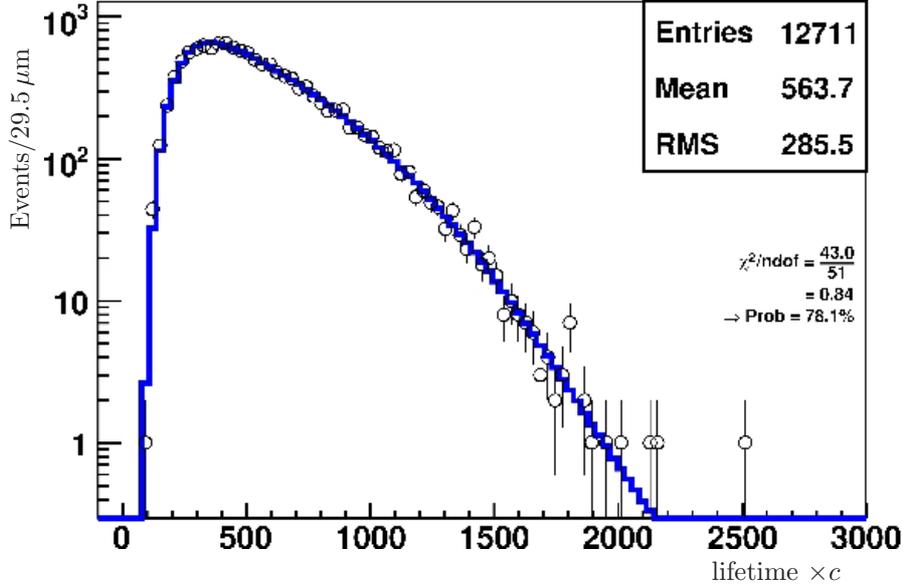
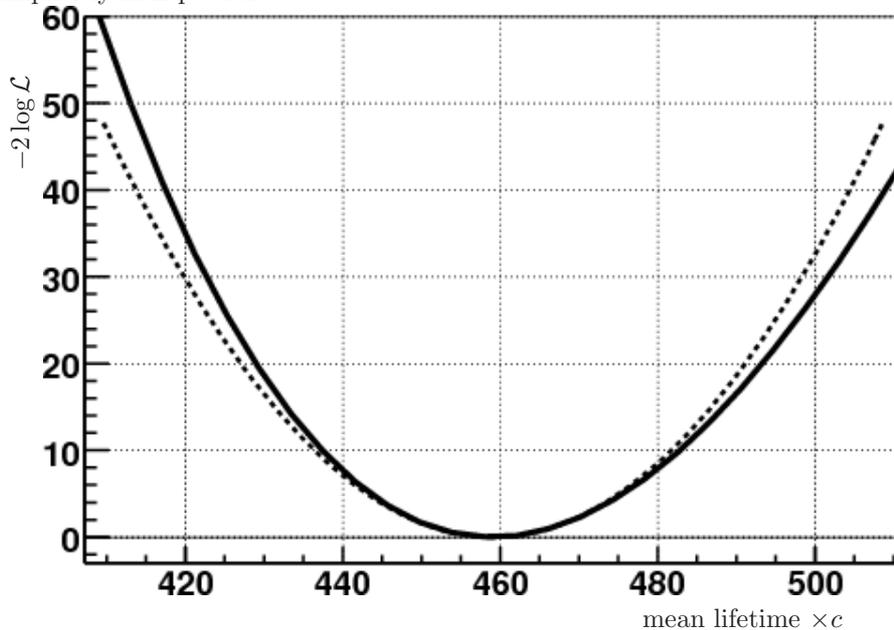
\begin{figure}
\caption{ The solid line is the likelihood curve for the fit to $\sim
 13k$ MC-generated signal events. The broken line represents the
 parabolic approximation made implicitly in
 \eqref{eq:varianceFromLogLik}.
 \label{fig:LL}
}
\input{Fig_ll.tex}
\end{figure}
 The result of the fit to $12,711$ signal events with an input lifetime of
 \un{c\tau=462}{\mu m}, is \un{c\tau = 459.1^{+7.3}_{-7.1}}{\mu
 m}. The agreement between the fit and the MC data is excellent, as
 can be seen in \figref{fig:fit}.

 The average statistical power per event, calculated using
 \eqref{eq:defStatPower}, is \un{31}{\%}. Using this in
 \Eqref{eq:varianceCalculated} yields an error estimate, {\it ignoring
 measurement errors,} of \un{7.2}{\mu m}, in good agreement with the
 numerical error estimate from the full likelihood.
  The solid line in \figref{fig:LL} represents the
 $-2\log(\mathrm{likelihood})$ curve for this fit, normalised to
 $-2\log \mathcal{L}=0$ at its minimum. The broken line represents the
 parabolic ($-2 \log(\mathrm{Gaussian})$) approximation implicitly
 made in \eqref{eq:varianceFromLogLik}. The agreement between the
 parabolic approximation and the actual likelihood for our data sample
 of $~13k$ events is very good within the range relevant for
 calculating $1\sigma$ errors, i.e., for this normalisation, up to
 $(-2 \log \mathcal{L}) = 1$. The errors estimated using the full
 likelihood are $+7.3\mu m,\; -7.1\mu m$, in excellent agreement with
 the parabolic approximation of $\pm 7.2 \mu m$. The difference
 between the two curves increases for larger $(-2 \log\mathcal{L})$
 values, as the asymmetry in the lifetime errors becomes more
 pronounced, but it remains reasonable over the entire range of the
 plot, up to $(-2\log\mathcal{L}) \approx 40$ and beyond. This implies
 that the parabolic approximation remains reasonable even for much
 smaller data samples containing only a few hundred events.

 This gives us confidence that the formula derived in
 \eqref{eq:varianceCalculated} is correct, and that, for the purpose
 of estimating the statistical uncertainty of a lifetime measurement,
 the approximations we made are justified for B hadron lifetime
 measurements at CDF, where event samples are usually large and
 typical event-by-event lifetime errors are about \un{60}{fs}, small
 compared to B hadron lifetimes of about \un{\sim 1.5}{ps}.

\section{Summary}
\label{sec:summary}
 We quantified the statistical effect of upper lifetime cuts in
 lifetime measurements for the simplified case that the event-by-event
 lifetime errors are small compared to the lifetime to be measured.

 We found that the effect of an upper lifetime cut is generally much
 more dramatic than the mere loss of events would suggest. The
 greatest impact of such a cut is a reduction in the statistical
 significance of each event for the purpose of lifetime
 measurements. For example an upper lifetime cut at twice the mean
 lifetime to be measured loses only $14\%$ of events, but the
 statistical power per event is reduced by $72\%$. The combined effect
 is equivalent to a reduction in sample size by a factor $4$, thus
 doubling the statistical error.

 We verified our calculation using simulated \prt{B^0 \to \pi\pi}
 events at CDF. In the trigger scenario used as an example,
 the statistical power per event is reduced by \un{69\%} due to the
 upper impact parameter cuts applied by the trigger.

\section*{Acknowledgements}
 \label{sec:acknowledgments}
 Many thanks go to Farrukh Azfar, Nicola Pounder and Azizur Rahaman for
 providing the Monte Carlo sample; and my colleagues in the
 Oxford/Pittsburgh collaboration for fitting B lifetimes: Farrukh
 Azfar, Joe Boudreau, Todd Huffman, Louis Lyons, Sneha Malde, Nicola
 Pounder and Azizur Rahaman.

%% file: Fig_statPower.tex
\parbox{0.58\textwidth}{
\rotatebox{90}{\parbox{0.4\textwidth}{\mbox{} \hfill stat power per
  event $\mathcal{P}$}}
\includegraphics[width=0.57\textwidth]{%
\fig/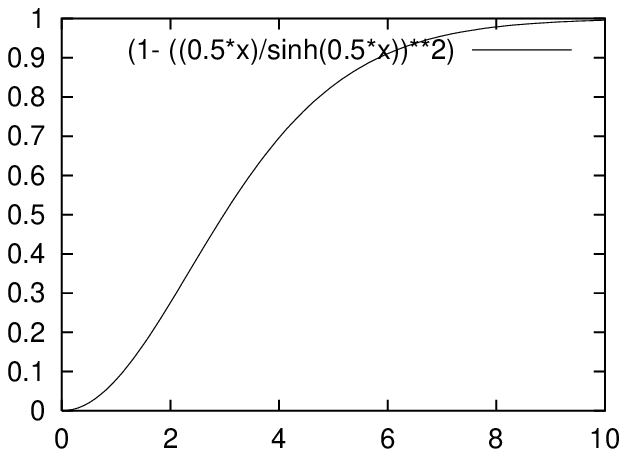}
\\\parbox{0.57\textwidth}{\mbox{}\hfill $\Delta  t / \tau$}
}

%% file: Fig_lifeFit.tex
\parbox{0.95\textwidth}{
\mbox{
\rotatebox{90}{\mbox{}\hspace{12em}Events/\un{29.5}{\mu m}}
\includegraphics[width=0.93\textwidth]{\fig/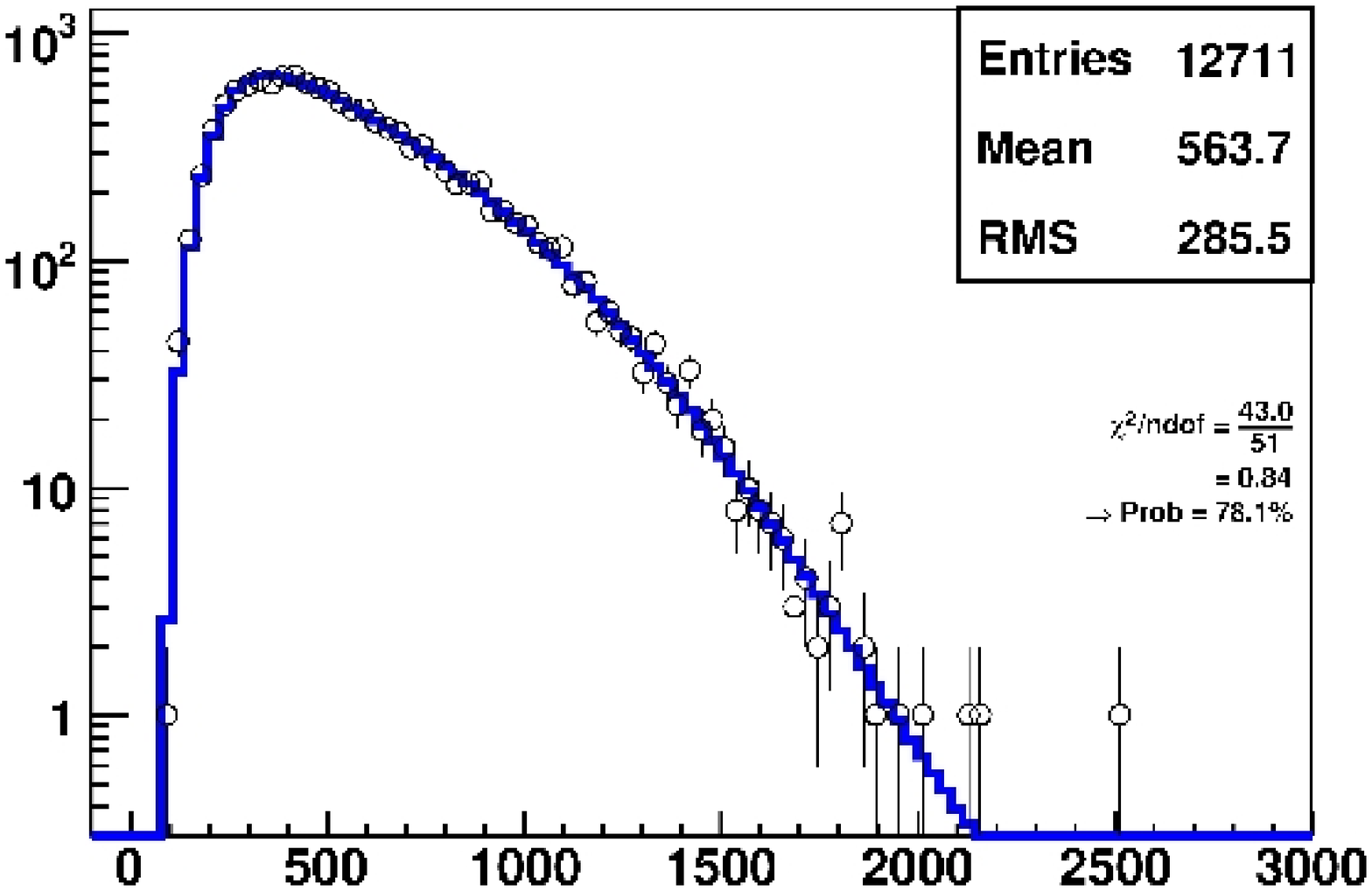}
}
\\\mbox{}\hfill lifetime $\times c$ \hspace{1em}\mbox{}
}

%% file: Fig_ll.tex
\parbox{0.95\textwidth}{
\mbox{
\rotatebox{90}{\mbox{}\hspace{16em}$-2\log\mathcal{L}$}
\includegraphics[width=0.93\textwidth]{\fig/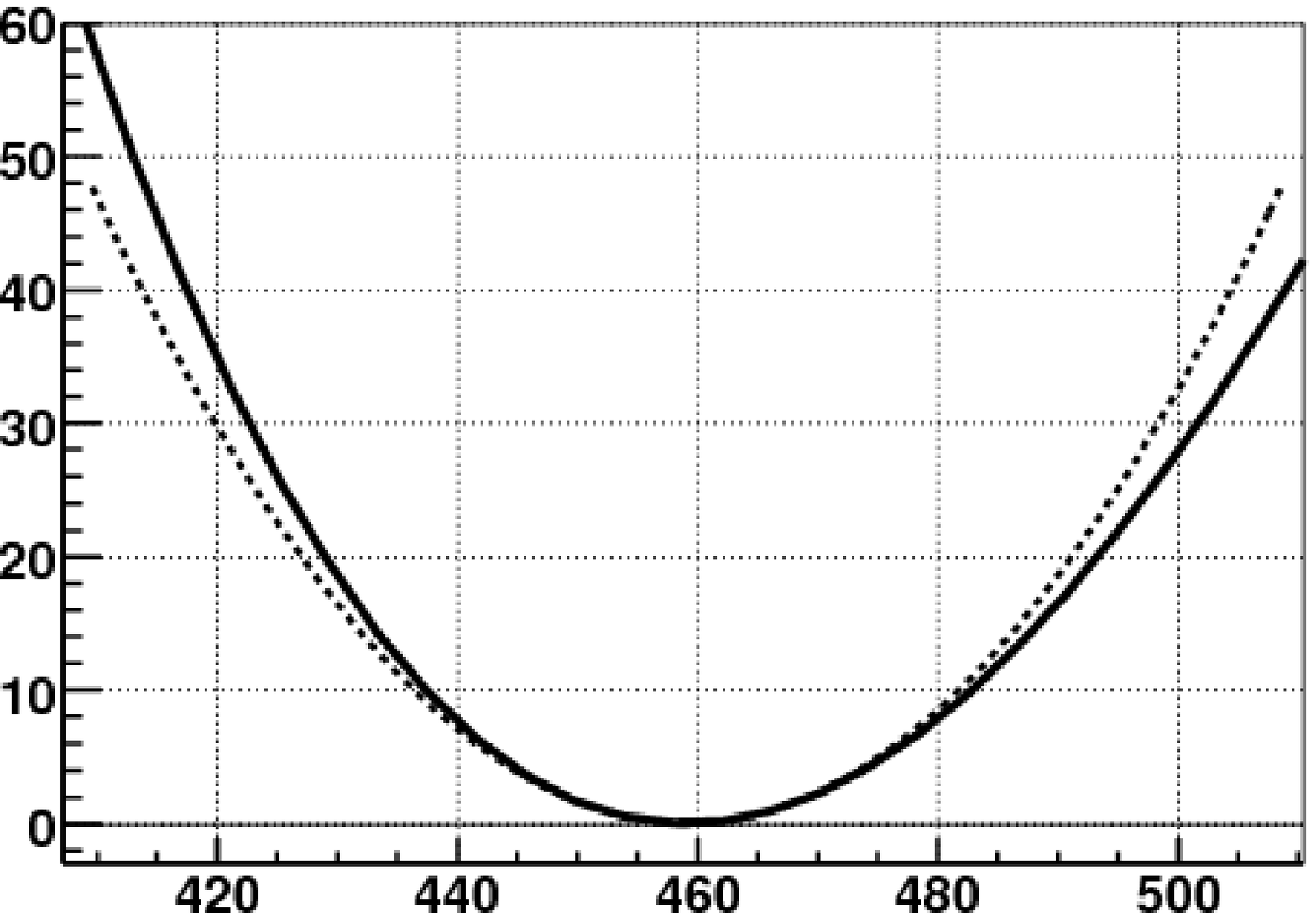}}
\\\mbox{}\hfill mean lifetime $\times c$ \hspace{1em}\mbox{}
}